%% file: main_old.tex
\documentclass[conference]{IEEEtran}
\IEEEoverridecommandlockouts
\usepackage{cite}
\usepackage{amsmath,amssymb,amsfonts}
\usepackage{algorithmic}
\usepackage{graphicx}
\usepackage{textcomp}
\usepackage{xcolor}
\usepackage{balance}
\def\BibTeX{{\rm B\kern-.05em{\sc i\kern-.025em b}\kern-.08em
    T\kern-.1667em\lower.7ex\hbox{E}\kern-.125emX}}
\begin{document}

\title{Summarizing CPU and GPU Design Trends with Product Data
}

\author{
\IEEEauthorblockN{
Yifan Sun, Nicolas Bohm Agostini, Shi Dong, and David Kaeli
}
\IEEEauthorblockA{
Northeastern University\\
Email: \{yifansun, agostini, shidong, kaeli\}@ece.neu.edu}
}

\maketitle


\input{tex/abstract.tex}
\input{tex/intro.tex}
\input{tex/methodology.tex}
\input{tex/findings.tex}
\balance
\input{tex/conclusion.tex}

\bibliographystyle{plain}
\bibliography{ref}

\end{document}

%% file: tex/abstract.tex
\begin{abstract}
Moore's Law and Dennard Scaling have guided the semiconductor industry for the past few decades. Recently, both laws have faced validity challenges as transistor sizes approach the practical limits of physics. We are interested in testing the validity of these laws and reflect on the reasons responsible. In this work, we collect data of more than 4000 publicly-available CPU and GPU products. We find that transistor scaling remains critical in keeping the laws valid. However, architectural solutions have become increasingly important and will play a larger role in the future. We observe that GPUs consistently deliver higher performance than CPUs. GPU performance continues to rise because of increases in GPU frequency, improvements in the thermal design power (TDP), and growth in die size. But we also see the ratio of GPU to CPU performance moving closer to parity, thanks to new SIMD extensions on CPUs and increased CPU core counts.  
 \end{abstract}

%% file: tex/intro.tex
\section{Introduction}

Moore's Law~\cite{schaller1997moore} and Dennard Scaling~\cite{dennard1974design} have guided the development of the semiconductor industry. Moore's Law predicted that the number of transistors on a chip would double every 18 months. Dennard scaling stated that the energy consumption of a chip would stay in proportion to the size of the chip. Shrinking transistor sizes (i.e., reducing the process size) would allow us to increase the computing capabilities of the device without consuming more energy. Moore's Law and Dennard Scaling charted a promising future for computers. As we reduce the transistor size, the performance of computing devices would exponentially increase over time.

Recently, the semiconductor industry has faced unprecedented challenges. As the transistor sizes is approaching the practical limitations of physics, reducing transistor sizes can hardly improve performance. To guide future semiconductor research, we need to revisit these historical development trends and find new directions that can drive performance improvement. 

The Central Processing Unit (CPU) and Graphic Processing Unit (GPU) are the two mainstream options for general-purpose computing. The former primarily serves as the processor for general tasks executed on mobile devices, personal computers, and servers. The latter takes on compute-intensive tasks for graphics rendering, big-data analytics, signal processing, artificial intelligence, and physics simulation~\cite{nickolls2010gpu}. To improve computing system performance, we need to better understand CPU and GPU design trends.

To study the trends, we collect and analyze the CPU and GPU data from public technical specifications of released products. Equipped with this data, we answer the following questions: 

\begin{itemize}
    \item Are Moore's Law and Dennard Scaling still valid? If so, what are the factors that keep the laws valid?
    \item Do GPUs still have computing power advantages over CPUs? Is the computing capability gap between CPUs and GPUs getting larger?
    \item What factors drive performance improvements in GPUs?
\end{itemize}

%% file: tex/methodology.tex
\section{Methodology}

\begin{figure*}[ht]
    \centering
    \includegraphics[width=\linewidth]{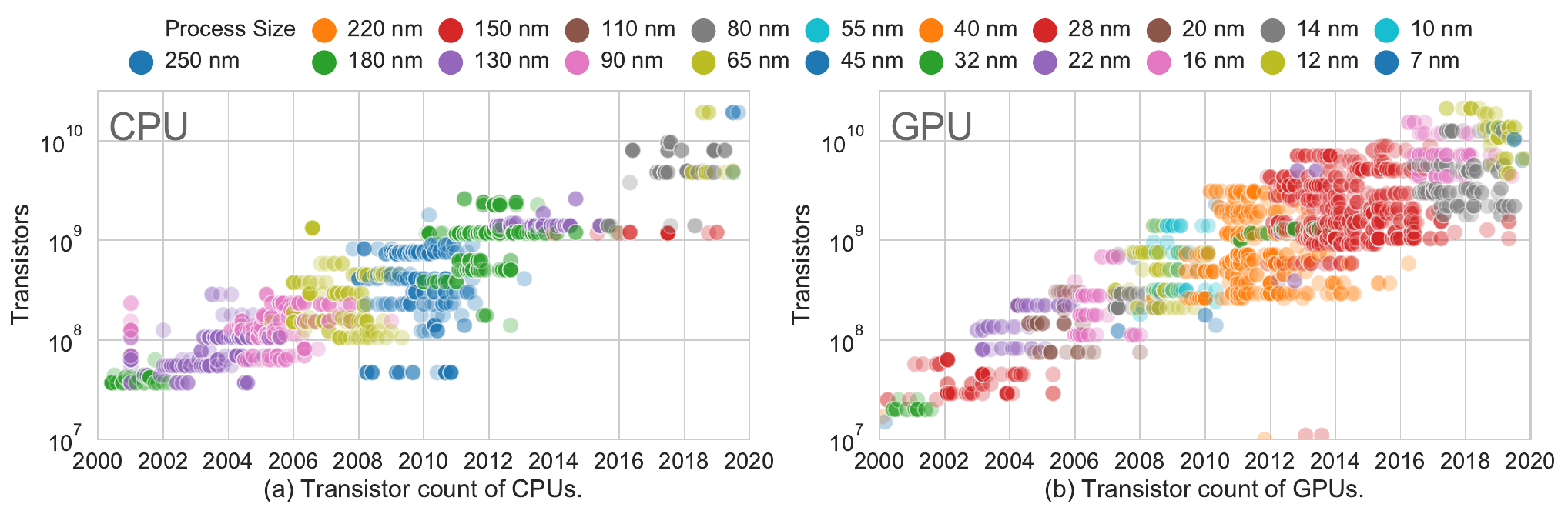}
    \caption{Moore's Law is still valid for both CPUs and GPUs.}
    \label{fig:moores_law}
\end{figure*}

We have collected data for all CPU and GPU products (to our best knowledge) that have been released by Intel, AMD (including the former ATI GPUs)\footnote{The data sources include www.techpowerup.com, wikichip.org, and vendor websites.}, and NVIDIA since January 1st, 2000. In total, the dataset consists of 4031 products (2102 CPUs and 1929 GPUs). For each product, we record its release date and properties, such as process size, die size, transistor count, base frequency and thermal design power~(TDP). 

A large number of GPU products (e.g., NVIDIA Tesla K80) have more than one GPU chip on a single PCB.  For these products, we only consider the properties of one of its chips. As an exception, the emerging Multi-Chip-Module (MCM) packaging technology effectively incorporates multiple chips in one package. We consider the multiple chips in a MCM package (e.g., AMD Threadripper CPUs) as a single chip. Because MCM packages are usually programmer transparent, a programmer can use multiple chips in an MCM package as if they are a single chip. The development of interposer technologies can also provide promising communication bandwidth between MCM chips. With recent technologies, MCM-CPUs and MCM-GPUs can be as efficient as single large chips~\cite{arunkumar2017mcm,tavana2019mcm}.

Another large group of products considered in this study represent integrated CPU-GPU devices. We consider such devices as CPUs since the CPUs are the dominating components in these devices. We only consider the CPU on an integrated CPU-GPU chip when calculating their theoretical computing capability. We use the Floating Point Operations Per Second (FLOPS) or Tera-FLOPS (TFLOPS) as the metrics to evaluate their theoretical single-precision and double-precision computing capabilities.

For device frequency, we only consider the base clocks. We do not consider the boost clocks as devices cannot continuously work at the boost clock frequencies. We estimate energy consumption using the Thermal Design Power (TDP). We admit that TDP may not be the best metric to evaluate a device's power consumption. Chips can easily consume more power than their TDP specification for a short periods of time. Chips may also maintain energy consumption under TDP by shutting down parts of the circuit. However, TDP can be a reasonable estimation of the average power consumption when a device runs regular workloads at its base clock.

Although we have most of the product data, we miss the critical data of CPUs produced by Intel since 2014. Intel stopped revealing its CPU transistor counts and the die sizes since the 7th generation Core CPU. Therefore, when presenting detailed studies that are related to transistor counts and die sizes (Figure.~\ref{fig:transistor_scaling}, \ref{fig:die_size}, \ref{fig:dennard_scaling}), we focus on the GPU data. Also, in these analyses, we only show devices that are manufactured by the Taiwan Semiconductor Manufacturing Company~(TSMC). We select TSMC because TSMC has produced the highest number of chips in our dataset. We only consider chips from a single foundry since different foundries tend to use different measurements when computing their process sizes.

%% file: tex/findings.tex
\section{Findings}

\begin{figure}[t]
    \centering
    \includegraphics[width=\linewidth]{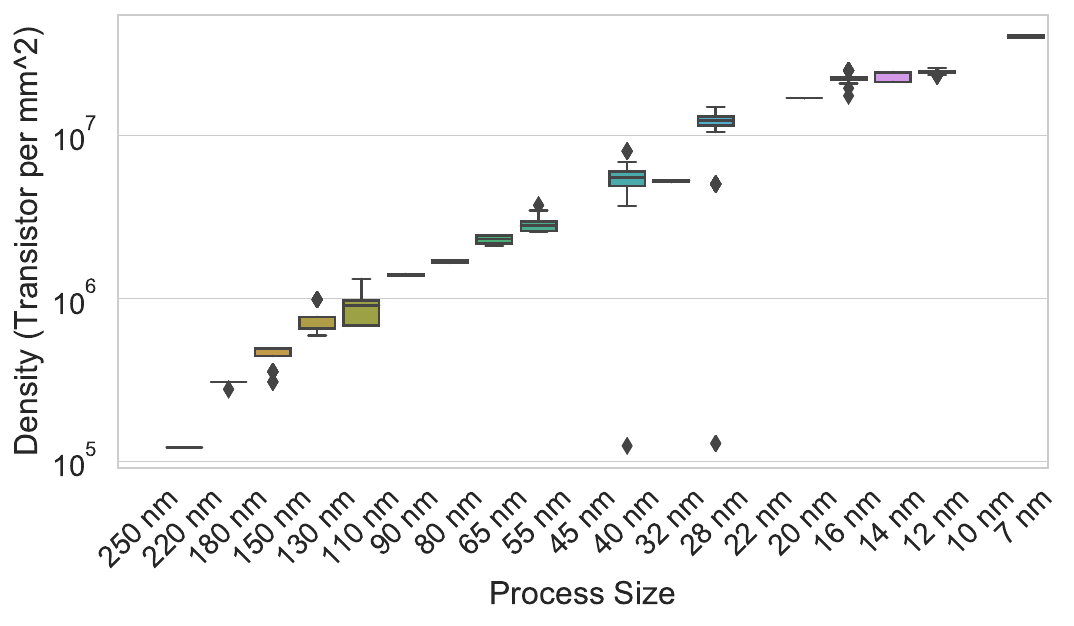}
    \caption{Transistor Scaling.}
    \label{fig:transistor_scaling}
\end{figure}

\begin{figure}[t]
    \centering
    \includegraphics[width=\linewidth]{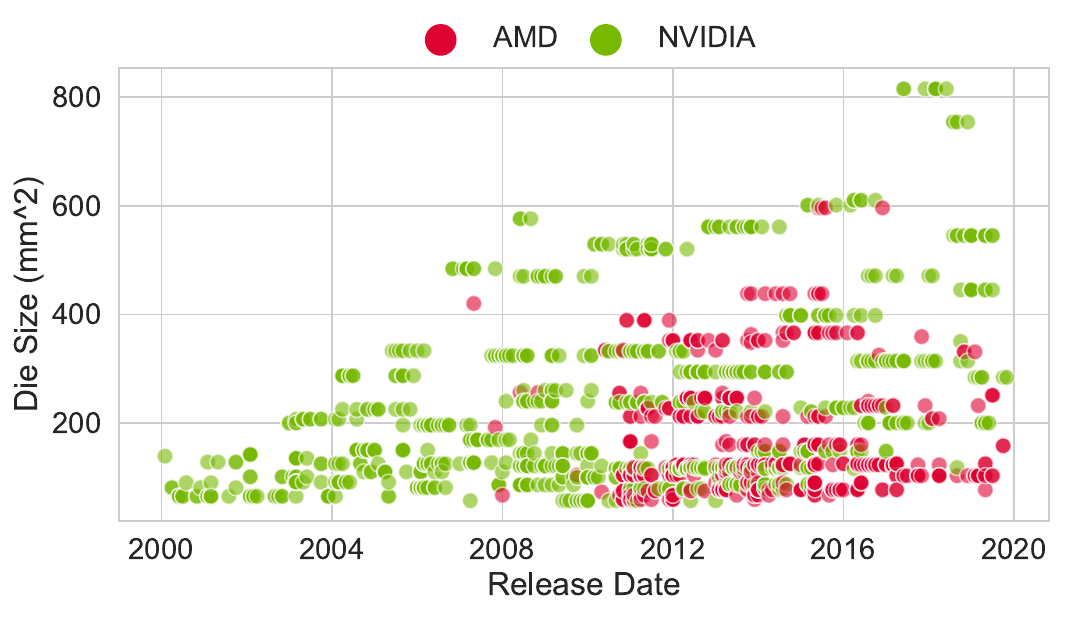}
    \caption{GPU Die Sizes.}
    \label{fig:die_size}
\end{figure}

\subsection{Moore's Law}

Researchers have been predicting the end of Moore's Law for a long time~\cite{mann2000end, kish2002end, waldrop2016chips, theis2017end}. However, Figure.~\ref{fig:moores_law} suggests the number of transistors on a chip is still increasing exponentially over time. For CPUs, we lack critical data from the years 2014 to 2017 because Intel stopped releasing transistor count and die size data since their 7th generation Core CPU. However, AMD restarted to produce high-end CPUs with large die-size recently. We can observe that the CPU transistor scaling trend is continuing to follow the pre-2014 trend. Also, Figure.~\ref{fig:moores_law} suggests that vendors tend to use new CMOS technologies in high-end products first. Low-end products may continue to use an older version of the CMOS process when the high-end products switch to a new technology node.

\begin{figure*}[t]
    \centering
    \includegraphics[width=\linewidth]{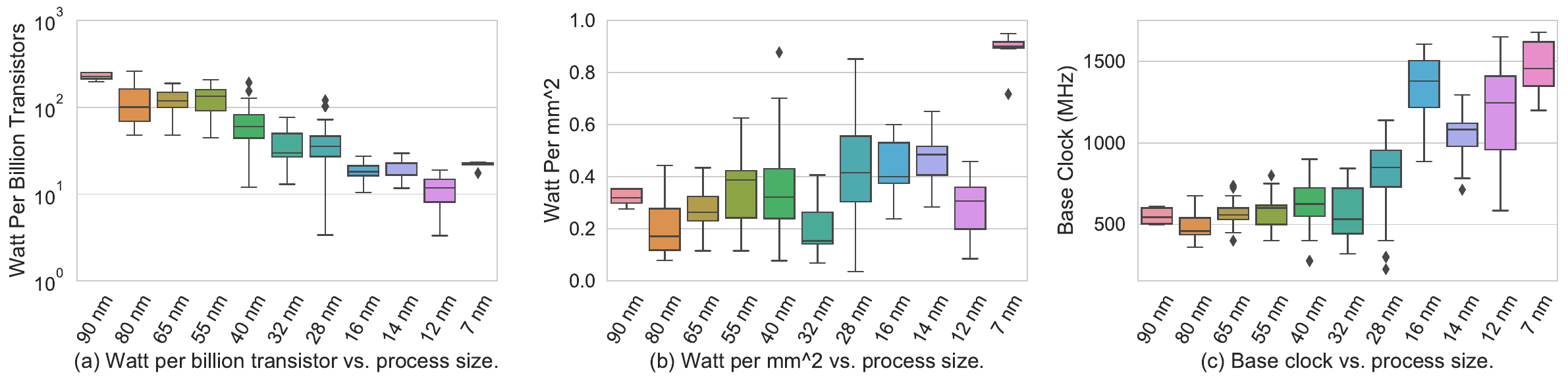}
    \caption{Process Size vs. (a) Energy consumption per billion transistor. (b) Energy consumption per area. (c) Base clock speed.}
    \label{fig:dennard_scaling}
\end{figure*}

The next question we need to ask is, "What drives increases in transistor count?" In Figure.~\ref{fig:transistor_scaling}, we see that transistor scaling is playing an essential role in increasing the die density (number of transistors per area). For the devices produced within a specific process size, the variance in terms of transistor density is small, even when considering GPUs from both AMD and NVIDIA. 

We also find that transistor scaling alone cannot keep up with Moore's Law. In Figure.~\ref{fig:die_size}, we observe a trend that the largest GPU die sizes are increasing over time. Increasing die sizes serve as the second factor pushing the overall increase in transistor counts for GPU devices.

\subsection{Dennard Scaling}

\begin{figure}[t]
    \centering
    \includegraphics[width=\linewidth]{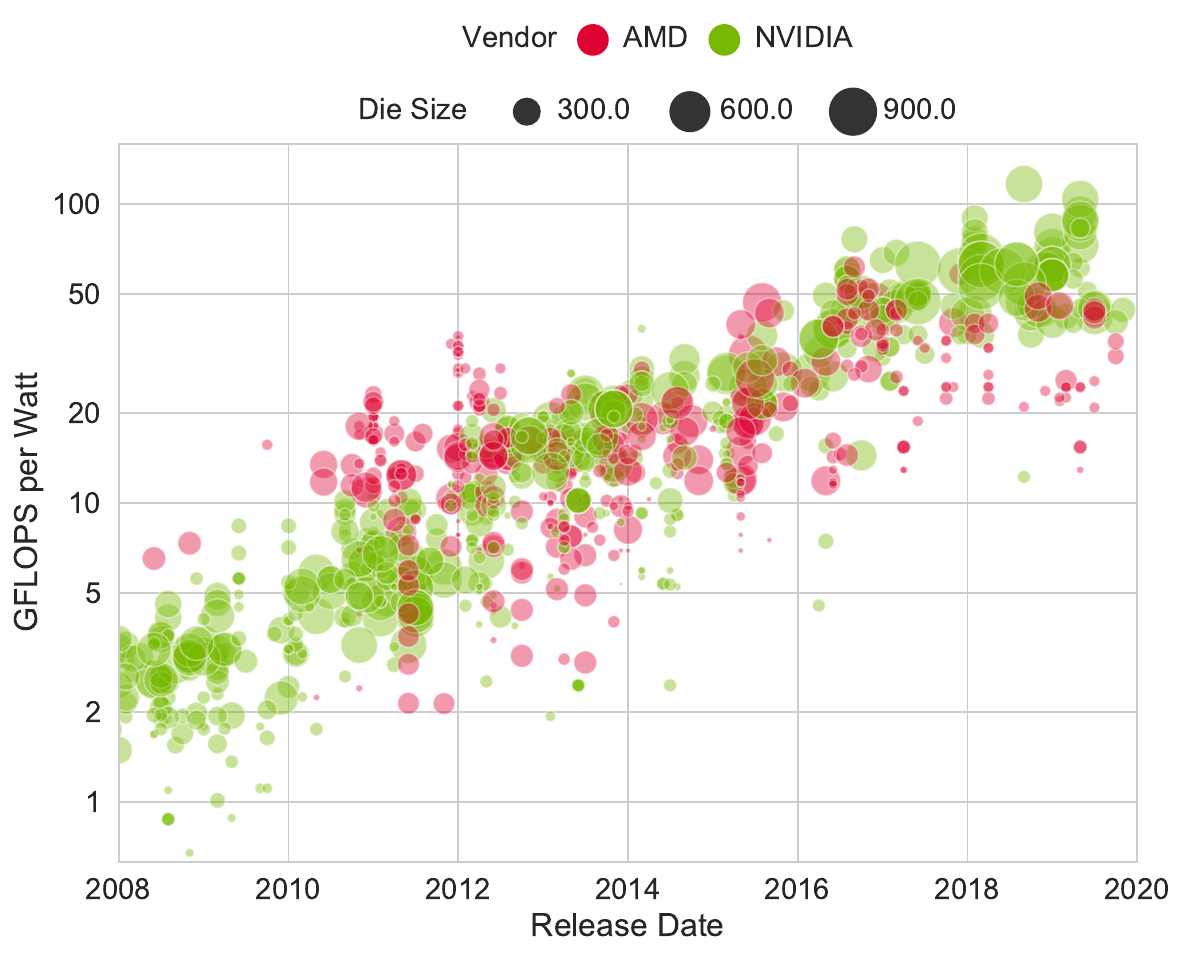}
    \caption{GPU FLOPS per watt.}
    \label{fig:koomeys_law}
\end{figure}

For a long time, Dennard Scaling has been another vital principle that guides the development of the semiconductor industry. According to Figure.~\ref{fig:dennard_scaling}, the general trend of Dennard Scaling was valid until the most recent technology nodes. The energy consumption per transistor keeps reducing with shrinking process sizes (Figure.~\ref{fig:dennard_scaling} (a)). The energy consumption per $mm^2$ also generally maintains a constant value between 1.2 $W/mm^2$ to 1.8 $W/mm^2$ in most of the technology nodes. The 32nm, 16nm, and 12nm chips are especially efficient in terms of watts per transistor or watts per $mm^2$.

We also see that Dennard Scaling is under pressure. The $W/mm^2$ is higher in the 28nm, 16nm, and 14nm nodes than in other nodes. The most recent 7nm technology has an unusually high energy density, exceeding the 98 percentile of the energy density of all other nodes. It is still too early to conclude the end of Dennard Scaling at the 7nm node since there have been only eight 7nm GPUs produced at the time of this report. We expect chip manufacturers to develop solutions to improve energy efficiency.

Despite the challenges to continue Dennard Scaling, frequency continues to increase with shrinking process sizes (as shown in Figure.~\ref{fig:dennard_scaling} (c)). This conflict suggests that shrinking transistor sizes impacts our ability to maintain Dennard Scaling. Architectural designs, such as dynamic voltage-frequency scaling (DVFS) and clock gating, have been playing an increasingly important role in maintaining the trend of Dennard Scaling.

Combining the GPU performance and energy consumption, we can see (Figure.~\ref{fig:koomeys_law}) that GPU energy efficiency is exponentially increasing. The FLOPS per watt doubles around every three to four years. 

\subsection{CPUs vs. GPUs}

\begin{figure*}[tb]
    \centering
    \includegraphics[width=\linewidth]{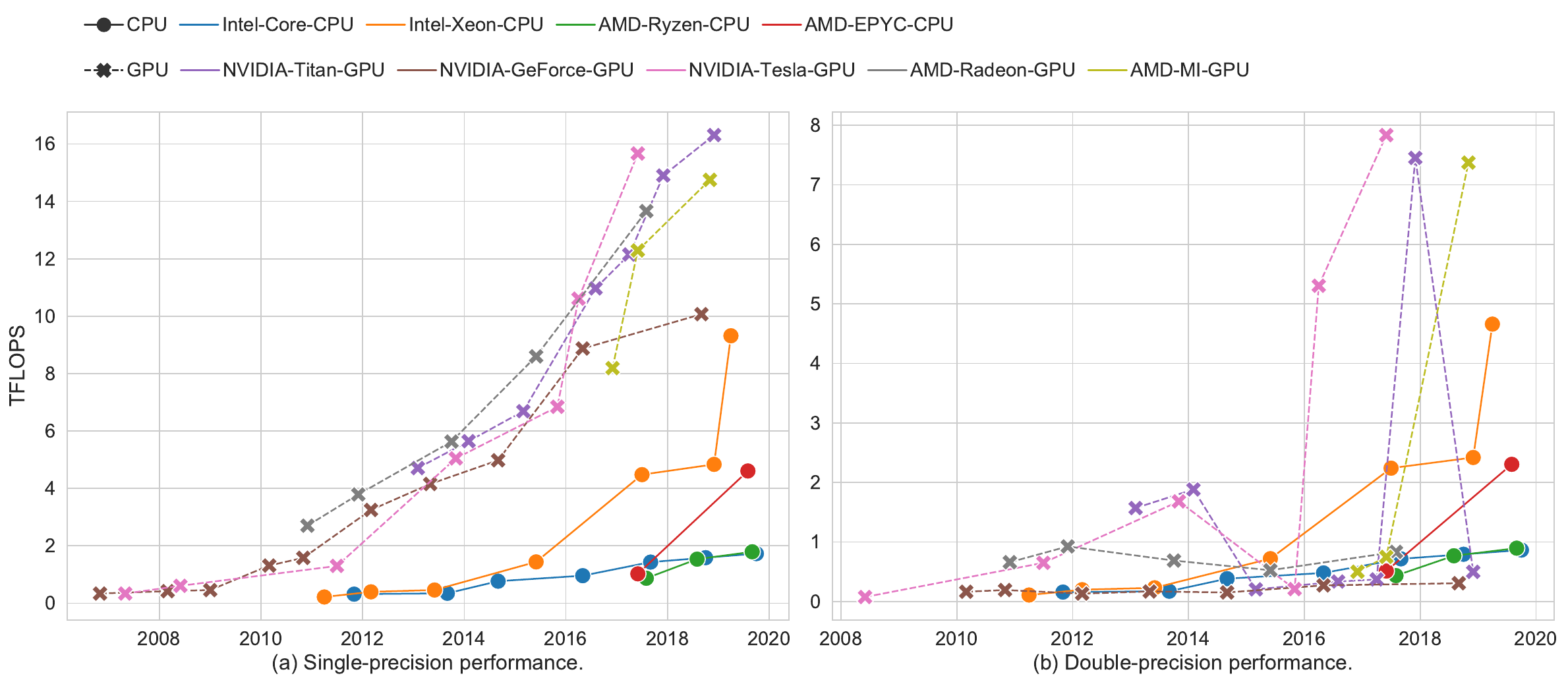}

    \caption[lol]{Comparing single-precision and double-precision performance of CPUs and GPUs.}   
    \label{fig:flops-date-cpu-gpu}
\end{figure*}

\begin{figure*}[t]
    \centering
    \includegraphics[width=\linewidth]{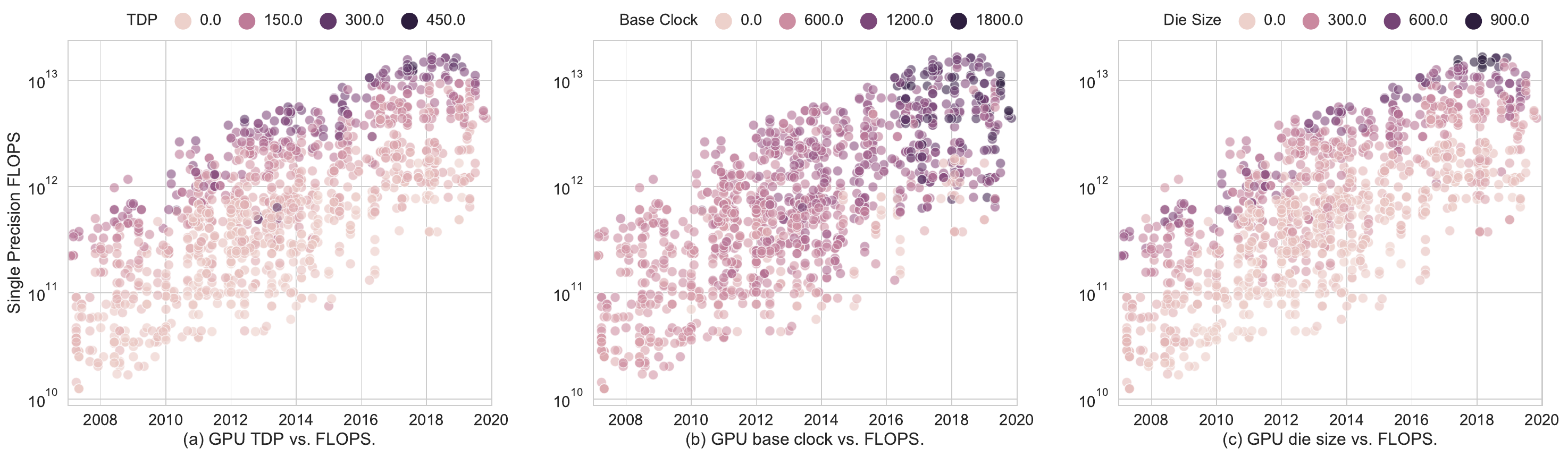}
    \caption{The trend of GPU Performance, TDP, base clocks, and die sizes.}
    \label{fig:gpu_perf}
\end{figure*}

Next, we compare the performance of CPUs and GPUs, as shown in Figure.~\ref{fig:flops-date-cpu-gpu}. Note that this analysis only considers the theoretical computing capabilities. The computing capabilities assumes the computing power that devices can deliver if all the computing resources are utilized. However, in real applications, always utilizing all the resources is generally impossible, and resource utilization depends on both the hardware and software implementation.

In general, GPUs still maintain significant advantages over consumer-level CPUs in terms of single-precision performance. The most powerful GPUs can deliver single-precision performance as high as 16 TFLOPS. A single GPU of today can be included in the Top-500 supercomputer list in 2008, which is only a decade ago. A CPU's single-precision computing capability is generally an order of magnitude lower than a GPU's. 

However, recent developments in CPU designs show a significant improvement in computing capabilities, reducing the performance gap between CPUs and GPUs. A server-grade CPU (Intel Xeon Platinum 9282) can deliver a theoretical computing power close to a state-of-the-art high-end gaming GPU (NVIDIA GeForce RTX 2080). It can also deliver more than half of the computing power of the most powerful GPU (an NVIDIA Tesla V100). The increase in CPU computing power is a combined result of recently developed CPU SIMD instructions extensions (e.g., SSE and AVX), an increased number of cores, and an increased CPU frequency. The reduction in the CPU-GPU performance gap suggests that we should not ignore the CPU computing power when considering CPU-GPU heterogeneous computing applications~\cite{heteromark,chai}. 

GPUs tend to have very different performance on double-precision computing. 
A few GPUs (e.g., the NVIDIA Tesla V100) have much higher performance than CPUs, while other 
GPUs struggle to be faster than CPUs in double-precision computing. GPUs have very different double-precision computing capabilities because GPU vendors design GPUs for different markets. For gaming-oriented GPUs, the ratio between the number of single-precision units and double-precision units is usually 32:1. For high-performance computing-oriented GPUs, the ratio is usually 2:1. Interestingly, for GPUs of the same product series, the ratio is not fixed. In recent years, with the development of new CPU technologies, CPUs can have a much higher double-precision performance than many GPUs. We suggest that users check the specifications of their CPUs and GPUs before using GPUs for double-precision computing.

\subsection{GPU Performance}

GPU performance is rapidly increasing. In the year 2019, the GPU with the smallest die size and consumed the least amount of energy has higher performance than the flagship GPU of 2007. Here, we analyze the factors that drive performance improvements of GPUs. 

According to Figure.~\ref{fig:gpu_perf} (a), the high-end GPUs always consume more energy compared to lower-end GPUs released around the same time. Over time, the TDP of flagship GPUs are increasing from around 150W to around 300W. High-TDP GPUs ($\approx$300W) start to appear around 2010-2012. Since then, the increases in TDP have slowed as the TDP approaches thermal dissipation limits imposed by current air or water cooling solutions. The high-end GPUs commonly have a 300W TDP. Only 3 AMD GPUs exceed the 300W TDP. 

We can also observe in Figure.~\ref{fig:gpu_perf} (a) that the power required to drive the same TFLOPS performance halves every three to four years. Newer devices have been consistently able to deliver more performance within the same power budget every year. If this trend continues, in 2020, we will see devices that consume less then 200W delivering more than 10 TFLOPS in performance.

The frequency increase is the most important factor that drives the performance improvement of GPUs. The frequency of GPUs tripled from around 600MHz to 1.8GHz (Figure.~\ref{fig:gpu_perf} (b)). The biggest jump happened around the year 2016 when NVIDIA released its Pascal GPUs. To be specific, the GTX-980 only runs at 1,216MHz frequency, while the GTX-1080 runs at a frequency of 1,733MHz. 

Another trend in GPU frequency is that high-end and low-end GPUs released each year tend to use similar frequencies. According to Figure.~\ref{fig:gpu_perf} (b), 
the frequency across devices targeting different market segments varies based on the time when the GPU is released.

Finally, as observed in Figure.~\ref{fig:gpu_perf} (c), the GPU die size is also increasing. The increased die size allows more transistors to reside in a single GPU. Note that the increases are coupled together with the reduction in the process size. Increasing die sizes reduce yield (the percentage of total manufactured chips that can pass validation tests). As the yield decreases, large die sizes push up the cost and price of GPU chips. Since CPUs have already started to use MCM packaging technologies to mitigate this problem, we believe that a MCM-GPU package is also a natural solution for GPUs.

%% file: tex/conclusion.tex
\section{Conclusion}

In this paper, we collected data from more than 4000 CPU and GPU products. With this data, we draw observations about Moore's Law and Dennard Scaling, in terms of their general validity. Transistor scaling clearly plays an essential role in increasing the number of transistors on a single chip and reducing the energy consumption per transistor. Architectural solutions, such as MCM packaging, DVFS, and clock gating, have become increasingly important in maintaining the historic scaling trends.

We also compared the performance of CPUs and GPUs. GPUs can still deliver a much higher single-precision computing power than consumer-level CPUs. However, recent development in parallel CPU computing has challenged the GPU's dominance. The gap between a CPU's and a GPU's computing capabilities was becoming smaller.

Finally, we analyzed the factors that drove up the performance improvement of GPUs. Frequency increases are the most critical factor that improved GPU performance. Die size and TDP increases also contributed to GPU performance improvements. We concluded that the energy efficiency (FLOPS per Watt) of GPUs doubles approximately every three to four years. 